\documentclass[amsmath,amssymb,aps,twocolumn,prb,floatfix,showpacs]{revtex4-1}
\usepackage[dvips]{graphicx}
\usepackage{bm}

\newcommand{\eInf}{{\epsilon_\infty}}

\newcommand{\IU}{{\frac{Z^{*2}}{\eInf a_0^3}}}   

\begin{document}
\title{First-principles accurate total-energy surfaces for polar structural distortions of BaTiO$_3$, PbTiO$_3$, and SrTiO$_3$:
       consequences to structural transition temperatures}

\author{Takeshi Nishimatsu$^{1}$}
\author{Masaya Iwamoto$^{1}$}
\author{Yoshiyuki Kawazoe$^{1}$}
\author{Umesh V. Waghmare$^{2}$}

\affiliation{
$^{1}$Institute for Materials Research (IMR), Tohoku University, Sendai 980-8577, Japan\\
$^{2}$Theoretical Sciences Unit, Jawaharlal Nehru Centre for Advanced Scientific Research (JNCASR),
Jakkur, Bangalore, 560 064, India
}

\begin{abstract}
Specific forms of the exchange correlation energy functionals in first-principles density functional theory-based
calculations, such as the local density approximation (LDA) and generalized-gradient approximations (GGA),
give rise to structural lattice parameters with typical errors of $-2$~\% and 2~\%.
Due to a strong coupling between structure and polarization,
the order parameter of ferroelectric transitions, they result in large errors in estimation of temperature
dependent ferroelectric structural transition properties. Here, we employ a recently developed GGA functional of
Wu and Cohen [Phys. Rev. B, {\bf 73}, 235116 (2006)] and determine total-energy surfaces for zone-center distortions
of BaTiO$_3$, PbTiO$_3$, and SrTiO$_3$, and compare them with the ones obtained with calculations based on
standard LDA and GGA. Confirming that the Wu and Cohen functional allows better estimation of structural properties at 0~K,
we determine a new set of parameters defining the effective Hamiltonian for ferroelectric transition in BaTiO$_3$.
Using the new set of parameters,
we perform molecular-dynamics (MD) simulations
under effective pressures $p=0.0$~GPa, $p=-2.0$~GPa, and $p=-0.005T$~GPa.
The simulations under $p=-0.005T$~GPa, which is for simulating thermal expansion,
show a clear improvement in
the cubic to tetragonal transition temperature and $c/a$ parameter of its ferroelectric tetragonal phase,
while the description of transitions at lower
temperatures to orthorhombic and rhombohedral phases is marginally improved.
Our findings augur well for use of Wu-Cohen functional in studies of ferroelectrics at nano-scale, particularly in the
form of epitaxial films where the properties depend crucially on the lattice mismatch.
\end{abstract}



\maketitle

\section{Introduction}
It is well known that first-principles density functional theory based calculations
within the local density approximation (LDA) underestimate lattice constants slightly (1--2\%), and
consequently calculated double-well total-energy surfaces\cite{King-Smith:V:1994,Hashimoto:N:M:K:S:I:JJAP:43:p6785-6792:2004}
for ferroelectric structural distortions of $AB$O$_3$ perovskite-type ferroelectrics are shallower
giving the theoretical transitions temperatures
much lower than their observed values
($T_{\rm C}=403~{\rm K}=0.0347~{\rm eV}$ for BaTiO$_3$, for example)\cite{JOHNSON:APL:7:p221:1965}.
In Monte-Carlo (MC) simulations\cite{Zhong:V:R:PRB:v52:p6301:1995} and
molecular-dynamics (MD) simulations\cite{Nishimatsu:feram:PRB2008} of
BaTiO$_3$, a perovskite-type ferroelectric,
an effective Hamiltonian\cite{King-Smith:V:1994,Zhong:V:R:PRB:v52:p6301:1995}
constructed from LDA calculations was used. To overcome the limitation of underestimation
of lattice constant, these simulations were carried out with a negative pressure of $-5$~GPa.
Similarly, Monte Carlo simulations for PbTiO$_3$ were carried out
with an effective Hamiltonian\cite{Waghmare:R:1997PRB} constructed from LDA calculations at the experimental
lattice constant. However, both of these schemes resulted in underestimation of $T_{\rm C}$.

There are two main sources of errors in estimation of $T_{\rm C}$ in such simulations:
(a) neglect of anharmonic coupling between soft modes and higher energy phonons in
construction of effective Hamiltonian, and (b) those arising from underestimation of lattice constants
in DFT calculations. While the former was assessed to be small in earlier work\cite{RABE:W:1992} and
can be partly corrected using a T-dependent pressure to yield a correct thermal expansion,
a systematic investigation of the latter would be useful in planning and evaluating
future first-principles simulations of ferroelectrics.

To overcome the limitation of DFT calculations in estimation of structural parameters in ferroelectrics,
Wu and Cohen introduced a new flavor of generalized gradient approximation (GGA),
and obtained an excellent agreement between calculated and experimentally observed lattice
constants at zero Kelvin for PbTiO$_3$ and BaTiO$_3$~\cite{Wu:C:PRB:73:p235116:2006}.
Recent theoretical works\cite{Bilc:PhysRevB.77.165107,Wahl:PhysRevB.78.104116}
further strengthened that the Wu and Cohen functional gives acceptable structural properties
of $AB$O$_3$ ferroelectrics such as lattice constants, $c/a$ ratio, atomic displacements,
phonon frequencies, Born effective charges, etc.

Here, we use the Wu and Cohen GGA-functional and determine
possibly more realistic total-energy surfaces of polar distortions
of BaTiO$_3$, PbTiO$_3$, and SrTiO$_3$ than those from LDA calculations.
Further, we construct an effective Hamiltonian for BaTiO$_3$ with a set of parameters
determined from first-principles calculations based on Wu-Cohen functional, and
estimate the three of its transition temperatures, and temperature dependent
structural properties. Through comparison with transitions properties obtained with
the LDA-based effective Hamiltonian and from experiment, we evaluate the efficacy of Wu-Cohen
functional in determination of finite-temperature properties.

In Sec.~\ref{sec:Formalism}, we describe the formalism of methods used in computations, with
a focus on details of the procedure for determination of the set of parameters of
the effective Hamiltonian, which is slightly different from the earlier works\cite{Zhong:V:R:PRB:v52:p6301:1995,
Waghmare:R:1997PRB}.
In Sec.~\ref{sec:results}, we present a comparative analysis of
calculated total-energy surfaces of BaTiO$_3$, PbTiO$_3$, and SrTiO$_3$ and include results
of MD simulations of BaTiO$_3$, and finally summarize our work and conclusions in
Sec.~\ref{sec:summary}.

\section{Methods of calculation and formalism}
\label{sec:Formalism}
\subsection{First-principles methods}
All calculations are performed with ABINIT code\cite{Gonze:ABINIT.ComputMaterSci:2002,ABINIT_CPC_2009}.
Bloch wave functions of electrons are expanded in terms of
plane waves with a cut-off energy of 60~Hartree,
and are sampled on an
$8\!\times\! 8\!\times\! 8$ grid of $k$-points
in the first Brillouin zone.
We use different choices of exchange correlation energy functionals.
For LDA calculations, we use the one parametrized by Teter\cite{Goedecker:T:H:1996}
along with Teter's extended norm-conserving pseudopotentials\cite{Teter:Pseudopotential:1993}.
For GGA calculations, we use ``PBE''\cite{Perdew:B:E:PRL:77:p3865-3868:1996}
and ``Wu and Cohen''\cite{Wu:C:PRB:73:p235116:2006} functionals,
along with Rappe's optimized pseudopotentials\cite{RAPPE:R:K:J:PRB:41:p1227-1230:1990} generated with Opium code\cite{opium}
and compare their results.

\subsection{Total-energy surface}
In 1994, King-Smith and Vanderbilt studied the total-energy surface
for zone-center distortions of perovskite-type ferroelectric oxides $AB$O$_3$
at zero temperature
using first-principles calculations with ultrasoft-pseudopotentials and
a plane-wave basis set.~\cite{King-Smith:V:1994}
Starting from the centrosymmetric cubic perovskite structure,
and using the normalized $\Gamma_{15}$ soft-mode eigenvector
$\bm{\xi}_\alpha$ ($=\bm{\xi}_x = \bm{\xi}_y = \bm{\xi}_z$, due to the cubic symmetry) of the
interatomic force constant (IFC) matrix\cite{Gonze:Lee:PhysRevB:55:10355},
they define displacements $v_{\alpha}^{\tau}$
of atoms $\tau$ (=$A$, $B$, O$_{\rm I}$, O$_{\rm II}$, O$_{\rm III}$)
in the Cartesian directions $\alpha (=x,y,z)$ as
\begin{equation}
  \label{eq:Eigenvector}
\bm{v}_\alpha=
\left(
  \begin{array}{c}
    v_\alpha^A         \\
    v_\alpha^B         \\
    v_\alpha^{\rm O_{\rm I}} \\
    v_\alpha^{\rm O_{\rm II}} \\
    v_\alpha^{\rm O_{\rm III}}
  \end{array}
\right)
= u_\alpha\bm{\xi}_\alpha
= u_\alpha
  \left(
    \begin{array}{c}
      \xi^A_\alpha \\
      \xi^B_\alpha \\
      \xi^{\rm O_{\rm I}}_\alpha \\
      \xi^{\rm O_{\rm II}}_\alpha \\
      \xi^{\rm O_{\rm III}}_\alpha
    \end{array}
  \right)\ ,
\end{equation}
with the scalar soft-mode amplitude $u_\alpha$.
Under the condition that the strain components
$\eta_i$~($i=1,\ \cdots,\ 6$; Voigt notation; $\eta_1=e_{xx}$, $\eta_4=e_{yz}$)
minimize the total energy for each $\bm{u}=(u_x, u_y, u_z)$,
they expressed the total energy as
\begin{equation}
  \label{eq:King-Smith-and-Vanderbilt}
  E^{\rm tot}
  = E^{0}
  + \kappa u^2
  + \alpha ' u^4
  + \gamma ' (u_x^2 u_y^2 +
              u_y^2 u_z^2 +
              u_z^2 u_x^2)\ ,
\end{equation}
where $u^2 = u_x^2 + u_y^2 + u_z^2$,
$E^{0}$ is the total energy of the cubic structure,
$\kappa$ is half the eigenvalue of the $\Gamma_{15}$ soft mode,
and $\alpha '$ and $\gamma '$ are anharmonic coefficients including the contribution
of relaxation of strain through its coupling with
atomic displacements.
Similar analysis is used in constructions of effective Hamiltonian in
Refs. \onlinecite{Zhong:V:R:PRB:v52:p6301:1995,Waghmare:R:1997PRB}, which assumes that anharmonic
coupling between soft modes and other IR-active modes is vanishingly small.

\begin{figure}
  \centering
  \includegraphics[angle=-90,width=80mm]{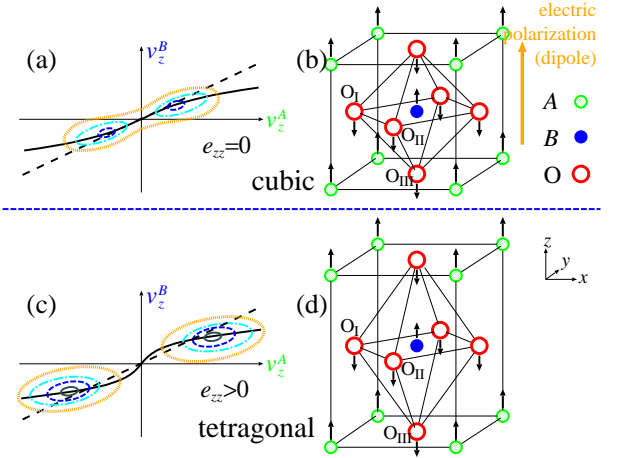}
  \caption{(Color online) Total-energy surfaces for zone-center distortions of a
    ferroelectric $AB$O$_3$ perovskite on a
    two-dimensional subspace $(v_z^A, v_z^B)$ of the
    atomic-displacement space $(v_z^A, v_z^B, v_z^{\rm O_{I}}, v_z^{\rm O_{II}}, v_z^{\rm O_{III}})$.
    (a) Schematic contour plot for atomic displacements from the centrosymmetric cubic structure (b), $e_{zz}=0$,
    is compared to (c) that from the tetragonal structure (d), $e_{zz}>0$.
    Thick solid lines are the valley lines for fixed $e_{zz}$'s.
    Dashed lines show the direction of the $\Gamma_{15}$ soft-mode eigenvector $\bm{\xi}$ at zero strain.
    Note that the direction is tangential to the valley line at $v_z^\tau=0$ for $e_{zz}=0$,
    but this is not the case for $e_{zz}\neq 0$.}
  \label{fig:contours}
\end{figure}

In 2004, Hashimoto, Nishimatsu {\it et al.} included some of these anharmonic effects by
redefining $u_\alpha$ as
\begin{equation}
  \label{eq:redefinition}
     u_{\alpha}=\sqrt{
                 \left( v_\alpha^A           \right)^2
                +\left( v_\alpha^B           \right)^2
                +\left( v_\alpha^{\rm O_{I}}   \right)^2
                +\left( v_\alpha^{\rm O_{II}}  \right)^2
                +\left( v_\alpha^{\rm O_{III}} \right)^2}\ ,
\end{equation}
and developed automatic computational method to determine
valley line of the total-energy surface in the 15-dimensional coordinate space of
atomic displacements\cite{Hashimoto:N:M:K:S:I:JJAP:43:p6785-6792:2004} (See Fig.~\ref{fig:contours}).
This method reveals that
normalized ``direction'' $\bm{\xi}$ of the atomic displacements
from the centrosymmetric cubic phase
to the distorted minima
is not constant as in Eq.~(\ref{eq:King-Smith-and-Vanderbilt}), but a function of $\bm{u}$, i.e. $\bm{\xi}(\bm{u})$,
and the total-energy surface generally cannot be
expressed with a 4th order function of
atomic displacement amplitude $\bm{u}$ as in Eq.~(\ref{eq:King-Smith-and-Vanderbilt}).
$\bm{\xi}(\bm{u})$ is determined by minimizing energy with respect to all atomic displacements
$\{ v^\tau_\alpha \}$ such that $\{ u_\alpha \}$ in Eq.~(3) is fixed.
In this paper,
we employ this valley line tracing method as implemented in the Patched ABINIT version 5.7.3
to determine accurate total-energy surfaces.
The patch file for ABINIT is in the EPAPS\cite{EPAPS2}.
The valley lines can be also calculated under any positive or negative pressure,
through use of enthalpy $H=E+pV$ and correspondingly
the enthalpy differences $H-H^0$ should be compared rather than total energy $E-E^0$.

\subsection{Effective Hamiltonian}
\label{sub:Effective:Hamiltonian}
The effective Hamiltonian constructed from first-principles calculations and
used in MD simulations is basically the same as that in Ref.~\onlinecite{Nishimatsu:feram:PRB2008,Waghmare:R:1997PRB},
\begin{multline}
  \label{eq:Effective:Hamiltonian}
  H^{\rm eff}
  = \frac{M^*_{\rm dipole}}{2} \sum_{\bm{R},\alpha}\dot{u}_\alpha^2(\bm{R})
  + \frac{M^*_{\rm acoustic}}{2}\sum_{\bm{R},\alpha}\dot{w}_\alpha^2(\bm{R})\\
  + V^{\rm self}(\{\bm{u}\})+V^{\rm dpl}(\{\bm{u}\})+V^{\rm short}(\{\bm{u}\})\\
  + V^{\rm elas,\,homo}(\eta_1,\cdots\!,\eta_6)+V^{\rm elas,\,inho}(\{\bm{w}\})\\
  + V^{\rm coup,\,homo}(\{\bm{u}\}, \eta_1,\cdots\!,\eta_6)+V^{\rm coup,\,inho}(\{\bm{u}\}, \{\bm{w}\})\\
  -Z^*\sum_{\bm{R}}\bm{\mathcal{E}}\!\cdot\!\bm{u}(\bm{R})~.
\end{multline}
Detailed explanation of symbols in the effective Hamiltonian
can be found in
Refs.~\onlinecite{King-Smith:V:1994},
\onlinecite{Zhong:V:R:PRB:v52:p6301:1995},
and \onlinecite{Nishimatsu:feram:PRB2008}.
We newly introduced 6th-order terms with coefficients $k_1$, $k_2$, and $k_3$
and an 8th-order term $k_4 u^8(\bm{R})$
to the local-mode self-energy $V^{\rm self}(\{\bm{u}\})$ as
\begin{multline}
  \label{eq:V:self}
  V^{\rm self}(\{\bm{u}\}) =
  \sum_{\bm{R}}
  \Bigl\{
  \kappa_2 u^2(\bm{R})
  + \alpha   u^4(\bm{R}) \\
  + \gamma \left[
                   u_y^2(\bm{R}) u_z^2(\bm{R})
                 + u_z^2(\bm{R}) u_x^2(\bm{R})
                 + u_x^2(\bm{R}) u_y^2(\bm{R})
           \right]\\
  + k_1 u^6(\bm{R})
  + k_2 \bigl[
                u_x^4(\bm{R}) ( u_y^2(\bm{R}) + u_z^2(\bm{R}))\\
              + u_y^4(\bm{R}) ( u_z^2(\bm{R}) + u_x^2(\bm{R}))
              + u_z^4(\bm{R}) ( u_x^2(\bm{R}) + u_y^2(\bm{R}))
        \bigr]\\
  + k_3 u_x^2(\bm{R}) u_y^2(\bm{R}) u_z^2(\bm{R}) + k_4 u^8(\bm{R})
  \Bigr\}~,
\end{multline}
where $u^2(\bm{R})
= u_x^2(\bm{R})
+ u_y^2(\bm{R})
+ u_z^2(\bm{R})$.
We introduced this 6th-order terms to follow-up the total-energy surface precisely.
The 8th-order term is for preventing the $|\bm{u}|\to\infty$ breakdown under negative $k_1$.


In next sections \ref{subsec:B11} and \ref{subsec:RF},
we explain how to determine the parameters for
the effective Hamiltonian of Eq.~(\ref{eq:Effective:Hamiltonian}) in detail.

\subsection{Elastic coefficients and total-energy surface}
\label{subsec:B11}
Elastic constants expressed in energy unit
$B_{11}=a_0^3C_{11}$ and
$B_{12}=a_0^3C_{12}$,
where $a_0$ is the equilibrium lattice constant in cubic structure,
can be calculated by deforming the cubic unit cell of $AB$O$_3$ with strain tensors
\begin{equation}
  \label{eq:uniform}
  \overleftrightarrow{\epsilon} =
  \left(
    \begin{array}{ccc}
      \delta & 0 & 0 \\
      0 & \delta & 0 \\
      0 & 0 & \delta \\
    \end{array}
  \right)
\end{equation}
and
\begin{equation}
  \label{eq:elongate}
  \overleftrightarrow{\epsilon} =
  \left(
    \begin{array}{ccc}
      0 & 0 & 0 \\
      0 & 0 & 0 \\
      0 & 0 & \delta \\
    \end{array}
  \right)~.
\end{equation}
Deformation in Eq.~(\ref{eq:uniform}) alters the total energy from its equilibrium value $E^0$ by
\begin{equation}
  \label{eq:E-uniform}
  E(\delta)=E^0 + \frac{3}{2}(B_{11}+2B_{12})\delta^2 + O(\delta^4)~.
\end{equation}
More precisely,
volume dependence of total energy may be fitted with
the Birch-Murnaghan equation of state~\cite{MurnaghanProcNatAcadSciUSA,Birch:PhysRev.71.809,OLAndersonTheUseOfUltrasonic}.
Deformation in Eq.~(\ref{eq:elongate}) gives
\begin{equation}
  \label{eq:E-elongate}
  E(\delta)=E^0 + \frac{1}{2}B_{11}\delta^2 + O(\delta^4)~.
\end{equation}
For $B_{44}=a_0^3C_{44}$, deformation
\begin{equation}
  \label{eq:44}
  \overleftrightarrow{\epsilon} =
  \left(
    \begin{array}{ccc}
      0 & \delta & \delta \\
      \delta & 0 & \delta \\
      \delta & \delta & 0 \\
    \end{array}
  \right)
\end{equation}
and
\begin{equation}
  \label{eq:E-44}
  E(\delta)=E^0 + \frac{3}{2}B_{44}\delta^2 + O(\delta^4)
\end{equation}
can be used.

$B_{1xx}$, $B_{1yy}$, and $B_{4yz}$, the coupling coefficients defined in Ref.~\onlinecite{King-Smith:V:1994},
are determined from quadratic $u$ dependence of strain.
In the case of $[110]$ distortion (see Fig.~\ref{fig:comp-bto}(e), for example),
\begin{subequations}
\begin{align}
  \label{eq:110:a}
  e_{xx} = a_{xx}u^2\\
  \label{eq:110:b}
  e_{xy} = a_{xy}u^2\\
  \label{eq:110:c}
  e_{zz} = a_{zz}u^2
\end{align}
\end{subequations}
emerge
\begin{subequations}
\begin{align}
  \label{eq:B1xx}
  B_{1xx} &= -4 B_{11} a_{xx} + 2 (B_{11}-2B_{12}) a_{zz}\\
  \label{eq:B1yy}
  B_{1yy} &= -4 B_{12} a_{xx} - 2 B_{11} a_{zz}\\
  \label{eq:B4yz}
  B_{4yz} &= -2 B_{44} a_{xy}~.
\end{align}
\end{subequations}

Anharmonic coefficients in the on-site energy
$\alpha$, $\gamma$, $k_1$, $k_2$, $k_3$, and $k_4$ in Eq.~(\ref{eq:V:self}) are determined from $u$-dependences
of total energies of $[001]$, $[110]$, and $[111]$ distortions as
\begin{subequations}
\begin{align}
  \label{eq:E001}
  E_{001}(u) =& \kappa u^2 + \alpha' u^4 + k_1 u^6 + k_4 u^8, \\
  \label{eq:E110}
  E_{110}(u) =& \kappa u^2 + (\alpha' + \frac{1}{4}\gamma') u^4 + (k_1 + \frac{1}{4}k_2)u^6 + k_4 u^8,\\
  \nonumber
  E_{111}(u) =& \kappa u^2 + (\alpha' + \frac{1}{3}\gamma') u^4 \\
  \label{eq:E111}
             & + (k_1 + \frac{2}{9}k_2 + \frac{1}{27}k_3)u^6 + k_4 u^8~.
\end{align}
\end{subequations}
With Eq.~(19a) and (19b) in Ref.~\onlinecite{King-Smith:V:1994},
$\alpha'$ and $\gamma'$ can be converted into
$\alpha$  and $\gamma$.
It should be mentioned that
it is quite difficult to express the total-energy surfaces
even with up to 8th order polynomial in wide range of $u$.
Therefore, we fit Eqs.~(\ref{eq:E001})--(\ref{eq:E111}) only
to the calculated data points within narrow range of $u$, e.g. $|u|\le 0.3$~[\AA] for BaTiO$_3$.

\subsection{Response-function calculations}
\label{subsec:RF}
We perform some response-function (RF) calculations\cite{Gonze:Lee:PhysRevB:55:10355} with ABINIT,
determine IFC matrices at the
$\bm{k}$-points of $\Gamma$, X, M, R, and center of the $\Sigma$ axis (See Fig.~\ref{fig:half}(A)),
then calculate their eigenvalues and eigenvectors.

We can determine local and short-range interaction parameters $\kappa_2$ and $j_1, \cdots, j_7$
in Ref.~\onlinecite{Zhong:V:R:PRB:v52:p6301:1995}
from selected eigenvalues $2\kappa(\Gamma_{\rm TO})$, $2\kappa({\rm X}_1)$, $2\kappa({\rm X}_5)$,
$2\kappa({\rm M}_{3'})$, $2\kappa({\rm M}_{5'})$, $2\kappa({\rm R}_{25'})$, and $2\kappa(\Sigma_{\rm LO})$.
Here, it is emphasised that $\kappa(\bm{k}_i)$ is
{\it half} of the mode-$i$ eigenvalue $2\kappa(\bm{k}_i)$ of the IFC matrix at each $\bm{k}$-point.
Practically,
$\kappa_2$ and $j_1, \cdots, j_7$ are determined
by solving linear equation
as described in Ref.~\onlinecite{Zhong:V:R:PRB:v52:p6301:1995}, in CGS units,
\begin{widetext}
\begin{subequations}
\begin{align}
  \label{eq:kappa:a}
  \kappa(\Gamma_{\rm TO}) &=& -\frac{2}{3}\pi\IU & +\kappa_2 & +2j_1 & +j_2 & +4j_3 & +2j_4 &       & +4j_6 & \\
  \label{eq:kappa:b}
  \kappa({\rm X}_1)     &=&  4.84372       \IU & +\kappa_2 & +2j_1 & -j_2 & -4j_3 & +2j_4 &       & -4j_6 & \\
  \label{eq:kappa:c}
  \kappa({\rm X}_5)     &=& -2.42186       \IU & +\kappa_2 &       & +j_2 &       & -2j_4 &       & -4j_6 & \\
  \label{eq:kappa:d}
  \kappa({\rm M}_{3'})   &=& -2.67679       \IU & +\kappa_2 & -2j_1 & +j_2 & -4j_3 & +2j_4 &       & +4j_6 & \\
  \label{eq:kappa:e}
  \kappa({\rm M}_{5'})   &=&  1.33839       \IU & +\kappa_2 &       & -j_2 &       & -2j_4 &       & +4j_6 & \\
  \label{eq:kappa:f}
  \kappa({\rm R}_{25'})  &=&                 & ~~~~\kappa_2 & -2j_1 & -j_2 & +4j_3 & +2j_4 &       & -4j_6 & \\
  \label{eq:kappa:g}
  \kappa(\Sigma_{\rm LO}) &=&  2.93226       \IU & +\kappa_2 & + j_1 &      &       &       & -2j_5 &       & -4j_7~,
\end{align}
\end{subequations}
\end{widetext}
and
\begin{equation}
  \label{eq:j6j7}
  0 = j_6-j_7~.
\end{equation}
Moreover, we newly assume that $j_5=0$ and $j_7=0$.
With this assumption,
we can omit one RF calculation for the center of the $\Sigma$ axis,
and we do not use Eq.~(\ref{eq:kappa:g}) and Eq.~(\ref{eq:j6j7}).
We can employ this assumption because $j_5$ and $j_7$ do not affect low energy polarization modes.

There may be an inconsistency between $\kappa$ from Eqs.~(\ref{eq:E001})--(\ref{eq:E111})
and $\kappa(\Gamma_{\rm TO})$ in Eq.~(\ref{eq:kappa:a}).
To keep the total-energy surfaces unchanged,
$\kappa$ from Eqs.~(\ref{eq:E001})--(\ref{eq:E111}) should be adopted.
Therefore, We add the difference between $\kappa$ and $\kappa(\Gamma_{\rm TO})$ to $\kappa_2$ as
\begin{equation}
  \label{eq:kappa2correction}
  \kappa_2 \leftarrow \kappa_2 + \left[\kappa - \kappa(\Gamma_{\rm TO})\right]~.
\end{equation}
This correction can be employed, because correction of $\kappa_2$
just leads parallel elevation of dispersion, e.g. Fig.~\ref{fig:half}(B).

From the calculated normalized $\Gamma_{15}$ soft-mode eigenvector $\bm{\xi}_\alpha$,
we determine the Born effective charge $Z^*$ and the effective mass $M^*_{\rm dipole}$ of the soft mode.
For $Z^*$, we also use the calculated effective charge tensor $Z^{*\tau}_{zz}$ for each atom $\tau$,
\begin{equation}
  \label{eq:EffectiveChargeXi}
  Z^* = \sum_\tau \xi_z^\tau\ Z^{*\tau}_{zz}~.
\end{equation}
The effective mass will be
\begin{equation}
  \label{eq:EffectiveMassXi}
  M^*_{\rm dipole} = \sum_\tau \{\xi_z^\tau\}^2 M^\tau~,
\end{equation}
where $M^\tau$ is the mass of atom $\tau$.

The optical dielectric constant $\eInf$ can be also determined in RF calculations.

\subsection{Conditions of molecular-dynamics simulations}
\label{subsec:MD:conditions}
MD simulations for BaTiO$_3$
with the effective Hamiltonian of Eq.~(\ref{eq:Effective:Hamiltonian})
are performed with our original MD code {\tt feram} (\url{http://loto.sourceforge.net/feram/}).
Details of the code can be found in Ref.~\onlinecite{Nishimatsu:feram:PRB2008}.
Temperature is kept constant in each temperature step
in the canonical ensemble
using the Nos\'e-Poincar\'e thermostat.\cite{Bond:L:L:JComputPhys:151:p114-134:1999}
This simplectic thermostat is so efficient that we can set the
time step to $\Delta t=2$~fs.
In our present MD simulations,
we thermalize the system for 180,000 time steps,
after which we average the properties for 20,000 time steps.
We used a supercell of system size $L_x\times L_y\times L_z = 14 \times 14 \times 14$
and small temperature steps in heating-up ($+1$~K/step) and cooling-down
($-1$~K/step) simulations.
It should be noted that the larger supercell size
and the more rapid heating-up and cooling-down
result in the larger temperature hysteresis.
The initial configuration are generated randomly:
$\langle u_\alpha \rangle = 0.11~{\rm \AA}$ and
$\langle u_\alpha^2 \rangle - \langle u_\alpha \rangle^2=(0.02~{\rm \AA})^2$.
We have checked that there is no dependence of results of these simulations
on initial configurations.

\newpage
\section{Results and Discussion}
\label{sec:results}
\begin{figure}
  \centering
  \includegraphics[width=72mm]{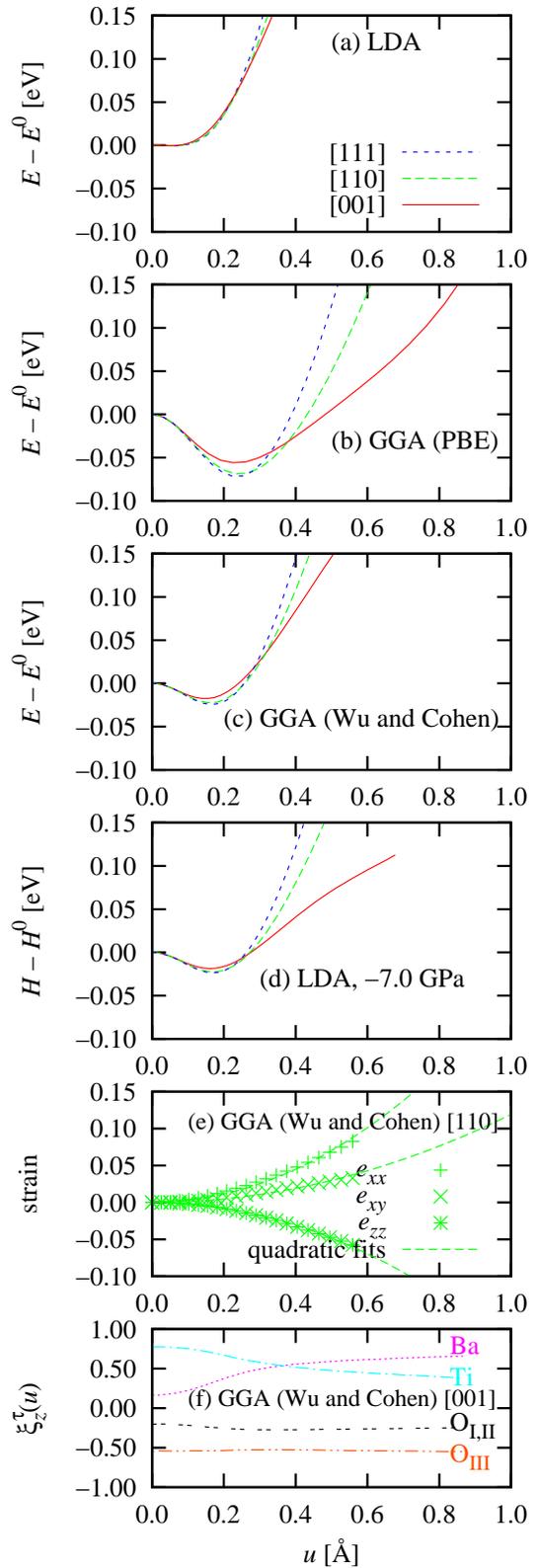}
  \caption{(Color online)
    (a)--(d) Total-energy surfaces for zone-center distortions of BaTiO$_3$.
    (e) $u$-dependence of strain along $[110]$ distortion.
    (f) ``Direction'' $\bm{\xi}(\bm{u})$ of atomic displacements along $[001]$ distortion.}
  \label{fig:comp-bto}
\end{figure}

Calculated total-energy curves
along $[001]$, $[110]$, and $[111]$ distortion directions of BaTiO$_3$
with three functionals
are shown in Fig.~\ref{fig:comp-bto}.
It is clear that LDA results in shallow double wells and
GGA~(PBE) results in rather deep wells, while
GGA~(Wu and Cohen) results in intermediate depths of the double well potentials.
Note that, in Fig.~\ref{fig:comp-bto}(a), double wells of LDA results cannot be recognized in this scale of energy.
It can be said that GGA~(Wu and Cohen) succeeds in reproducing total-energy surfaces at 0~K.
These results can be understood from estimated equilibrium cubic lattice constants
$a_0=$ 3.938, 4.034, and 3.986~\AA, respectively (See also Table~\ref{tab:a0}.).
GGA~(Wu and Cohen) results also reveal that
the ``direction'' $\bm{\xi}(\bm{u})$ of atomic displacements largely depends on $u$,
as shown in Fig.~\ref{fig:comp-bto}(f) even in BaTiO$_3$, which exhibits relatively small polar structural
distortions across its ferroelectric transition.
For emphasizing that the depth of the double wells
are strongly affected by equilibrium cubic lattice constant,
total-energy surfaces calculated with LDA
under negative pressure $-7.0$~GPa ($a_0=3.989$~\AA)
are calculated and show in Fig.~\ref{fig:comp-bto}(d).
This value of negative pressure, $-7.0$~GPa,
is selected to get similar total-energy surfaces as the GGA~(Wu and Cohen) results
from $-1.0, -2.0, \cdots, -9.0$~GPa calculations.
It can be seen that
LDA under certain negative pressure gives results similar to those of GGA~(Wu and Cohen).
\begin{table}
  \caption{Calculated equilibrium lattice constants $a_0$ for cubic phases
    are compared to experimental values observed just above $T_{\rm C}$.
  Experimental values are cited from
  Refs.~\onlinecite{Kay:Vousden:Philosophical.Magazine:40:p1019-1040:1949} and
       \onlinecite{JOHNSON:APL:7:p221:1965}                       for BaTiO$_3$ ($T_{\rm C}=403$~K),
  Ref.~\onlinecite{SHIRANE:H:S:PHYSICALREVIEW:80:p1105-1106:1950} for PbTiO$_3$ ($T_{\rm C}=763$~K), and
  Ref.~\onlinecite{Okazaki:K:MaterResBull:8:p545-550:1973}        for SrTiO$_3$ ($T_{\rm C}=106$~K).}
  \label{tab:a0}
  \centering
  \begin{tabular}{llll}
    \hline
    \hline
                & BaTiO$_3$ & PbTiO$_3$ & SrTiO$_3$ \\
    \hline
    experiment  & 4.010 \AA & 3.960 \AA & 3.896 \AA \\
    \hline
    LDA (Teter) & 3.938 \AA & 3.880 \AA & 3.845 \AA \\
    GGA (PBE)   & 4.034 \AA & 3.976 \AA & 3.946 \AA \\
    GGA (W\&C)  & 3.986 \AA & 3.930 \AA & 3.901 \AA \\
    \hline
    LDA under   & 3.989 \AA & 3.905 \AA & 3.899 \AA \\
    negative pressure
                & $-7.0$ GPa& $-4.0$ GPa& $-8.0$ GPa\\
    \hline
    \hline
  \end{tabular}
\end{table}

\begin{figure}
  \centering
  \includegraphics[width=80mm]{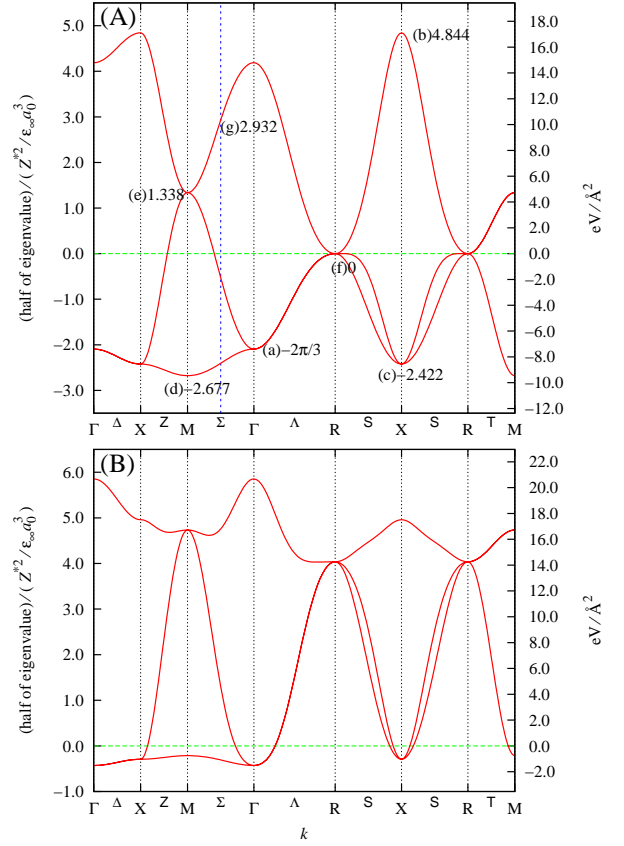}
  \caption{(Color online) (A) Half of eigenvalues of the $3 \times 3$ long-range dipole-dipole interaction
    matrix $\widetilde\Phi(\bm{k})$ (Fourier transform of Eq.~(10) in Ref.~\onlinecite{Nishimatsu:feram:PRB2008})
    are plotted along symmetric axes in the the first Brillouin zone of the simple-cubic lattice.
    Special points and $\bm{k}/(2\pi/a)=(\frac{1}{4}, \frac{1}{4}, 0)$ (the center of the $\Sigma$ axis) are
    indicated with vertical dotted lines. Labels (a)--(g)
    corresponds to Eqs.~(\ref{eq:kappa:a})--(\ref{eq:kappa:g}), respectively.
    Tics in the unit of $\IU$ is placed in left side.
    Tics in the unit of eV, in the case of the parameter set of Table~\ref{tab:parameters},
    is placed in right side.
    (B) Half of eigenvalues of the total (long-range $+$ short-range) interaction
    matrix $\widetilde\Phi^{\rm quad}(\bm{k})$ (Eq.~(13) in Ref.~\onlinecite{Nishimatsu:feram:PRB2008}).}
  \label{fig:half}
\end{figure}

From our calculations with GGA~(Wu and Cohen),
i.e. total-energy surfaces, $u$-dependence of strain (Fig.~\ref{fig:comp-bto}(e)),
IFC matrices, etc.,
we construct a new parameter set of effective Hamiltonian for BaTiO$_3$.
The parameters from Refs.~\onlinecite{King-Smith:V:1994} and \onlinecite{Zhong:V:R:PRB:v52:p6301:1995}
and those of present work are compared in Table~\ref{tab:parameters}.
\begin{table}
  \caption{Comparison of two set of parameters for the BaTiO$_3$ effective Hamiltonian.
    ``---'' indicates that values was not in use.
    ``n.a.'' indicates that values are not available.
    $p$ is constant or temperature $T$~[K] dependent effective negative pressures applied while MD simulations.
    $\kappa$ in Eqs.~(\ref{eq:E001})--(\ref{eq:E111}) and
    $\kappa(\bm{k}_i)$ in Eqs.~(\ref{eq:kappa:a})--(\ref{eq:kappa:g})
    are also listed.
    They are used to determine $\kappa_2$ and $j_1,\cdots,j_7$.
    $\xi_{z}^{\tau}$ are the soft-mode eigenvector.}
  \label{tab:parameters}
  \centering
  \begin{tabular}{rlrr}
    \hline
    \hline
             &              & Refs.    & present  \\
    \multicolumn{2}{c}{parameter}&\onlinecite{King-Smith:V:1994} and \onlinecite{Zhong:V:R:PRB:v52:p6301:1995}
                                             & work     \\
    \hline
    $p$      & [GPa]       & $-4.8$         & $-0.005T$    \\
    \hline
    $a_0$    & [\AA]       &  3.95           &  3.986        \\
    $B_{11}$  & [eV]        & 127.0           & 126.73        \\
    $B_{12}$  & [eV]        &  44.9           &  41.76       \\
    $B_{44}$  & [eV]        &  50.3           &  49.24       \\
    \hline
    $B_{1xx}$ & [eV/\AA$^2$] & $-211.$        &  $-185.35$   \\
    $B_{1yy}$ & [eV/\AA$^2$] & $-19.3$        &  $-3.2809$    \\
    $B_{4yz}$ & [eV/\AA$^2$] & $ -7.75$       &  $-14.550$    \\
    $\alpha$ & [eV/\AA$^4$] &   111.         &    78.99     \\
    $\gamma$ & [eV/\AA$^4$] & $-164.$       & $-115.48$     \\
    $k_1$    & [eV/\AA$^6$] & ---             & $-267.98$            \\
    $k_2$    & [eV/\AA$^6$] & ---             &   197.50            \\
    $k_3$    & [eV/\AA$^6$] & ---             &   830.20            \\
    $k_4$    & [eV/\AA$^8$] & ---             &   641.97            \\
    \hline
    $m^*$    & [amu]       & ---             &   38.24           \\
    $Z^*$    & [e]          &       9.956     &  10.33      \\
    $\epsilon_\infty$ &      &         5.24    &   6.87       \\
    $\kappa_2$&[eV/\AA$^2$] &         5.52    &   8.534           \\
    $j_1$    & [eV/\AA$^2$] &    $-2.657$      &         $-2.084$      \\
    $j_2$    & [eV/\AA$^2$] &      3.906       &         $-1.129$      \\
    $j_3$    & [eV/\AA$^2$] &      0.901       &           0.689       \\
    $j_4$    & [eV/\AA$^2$] &    $-0.792$      &         $-0.611$      \\
    $j_5$    & [eV/\AA$^2$] &      0.564       &           0.000       \\
    $j_6$    & [eV/\AA$^2$] &      0.360       &           0.277       \\
    $j_7$    & [eV/\AA$^2$] &      0.180       &           0.000       \\
    \hline
    \hline
    $\kappa$&[eV/\AA$^2$] &  $-1.695$      &   $-1.518$            \\
    \hline
    $\kappa(\Gamma_{\rm TO})$
            &[eV/\AA$^2$] &  n.a.         &    $-1.906$           \\
    $\kappa({\rm X}_1)$
            &[eV/\AA$^2$] &  n.a.          &     17.128           \\
    $\kappa({\rm X}_5)$
            &[eV/\AA$^2$] & n.a.           &    $-1.422$           \\
    $\kappa({\rm M}_{3'})$
            &[eV/\AA$^2$] & n.a.           &    $-1.143$           \\
    $\kappa({\rm M}_{5'})$
            &[eV/\AA$^2$] & n.a.           &     16.333           \\
    $\kappa({\rm R}_{25'})$
            &[eV/\AA$^2$] & n.a.           &     13.871            \\
    \hline
    \hline
    $\xi^A_z$ &             &   0.20          &           0.166      \\
    $\xi^B_z$ &             &   0.76          &           0.770      \\
    $\xi^{{\rm O}_{\rm I}}  _z$ & &$-0.21$        &     $-0.202$         \\
    $\xi^{{\rm O}_{\rm II}} _z$ & &$-0.21$         &    $-0.202$         \\
    $\xi^{{\rm O}_{\rm III}}_z$ & &$-0.53$         &    $-0.546$         \\
    \hline
    \hline
    $Z^{*A}_{zz}$ & [e]         & 2.75   & 2.741   \\
    $Z^{*B}_{zz}$ & [e]         & 7.16   & 7.492   \\
    $Z^{*{\rm O}_{\rm I}}  _{zz}$ &[e]& $-2.11$ &$-2.150$ \\
    $Z^{*{\rm O}_{\rm II}} _{zz}$ &[e]& $-2.11$ &$-2.150$ \\
    $Z^{*{\rm O}_{\rm III}}_{zz}$ &[e]& $-5.69$ &$-5.933$ \\
    \hline
    \hline
  \end{tabular}
\end{table}
As shown in Fig.~\ref{fig:half}(A),
without short-range interaction,
pure dipole-dipole long-range interaction results in
an antiferroelectric cell-doubling state as the most stable structure, corresponding
to the strongest instability at M point.
However, as shown in Fig.~\ref{fig:half}(B),
introduction of short-range interactions
$\kappa_2$ and $j_1, \cdots, j_7$ results in
the ferroelectric state as the most stable structure at the $\Gamma$ point.

\begin{figure}
  \centering
  \includegraphics[width=68mm]{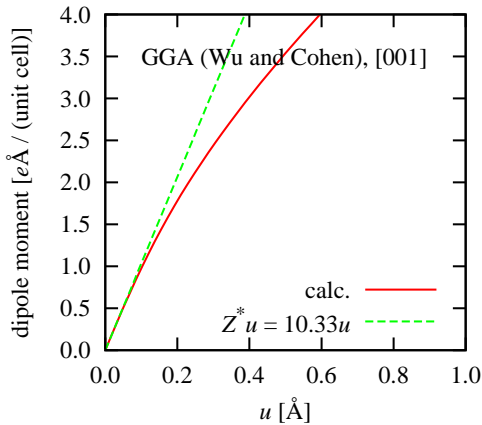}
  \caption{(Color online) Calculated dipole moment per unit cell as a function of $u$
    for atomic displacements along $[001]$ distortion (solid line).
    $Z^*u$ is also plotted for comparison (dashed line).}
  \label{fig:berry}
\end{figure}

In Fig.~\ref{fig:berry},
dipole moment per unit cell
as a function of $u$ for atomic displacements along $[001]$ distortion
calculated with the Berry-phase theory\cite{King-Smith:V:PRB:47:p1651-1654:1993}
is shown.
As compared with $Z^*u$, it can be seen that
linearity is broken at large $u$.
Since following MD simulations are based in energetics,
effects of this nonlinear $Z^*\!(\bm{u})$
mainly get folded into anharmonic terms,
i.e. $\alpha$, $\gamma$, $k_1$, $k_2$, $k_3$, and $k_4$.
However, still, there might be issues with intersite anharmonic interactions.
We leave the issues for future studies.

From the Table~\ref{tab:parameters}, it is evident that most parameters in the effective Hamiltonian
are sensitive to the choice of exchange-correlation functional, with the exception of
elastic constants and the mode effective charge. Largest change is seen in the parameters of
coupling between strain and polarization, as is expected from the fact that Wu-Cohen functional
gives a better estimate of lattice parameters and that they couple strongly with polarization.
Electronic contribution to the dielectric constant $\epsilon_\infty$ is further overestimated
with Wu-Cohen functional (it is typically 20 \% overestimated in an LDA-based calculation). Due to
the use of Hashimoto-Nishimatsu's valley tracing technique, description of the on-site potential
energy curve requires anharmonic terms expanded up to 8th order.

\begin{figure}
  \centering
  \includegraphics[width=55mm]{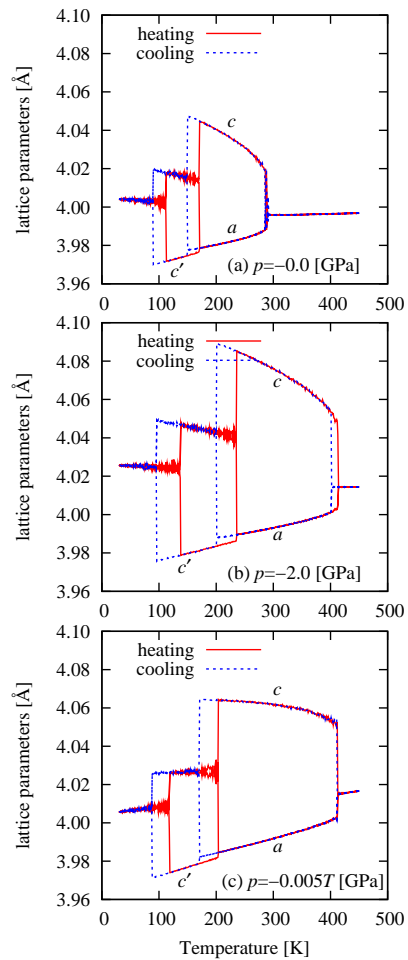}
  \caption{(Color online) Simulated temperature dependence of lattice parameters under
    (a)~$p=-0.0$~GPa, (b)~$p=-2.0$~GPa, (c)~$p=-0.005T$~GPa.}
  \label{fig:conpare-p}
\end{figure}

Using the Wu-Cohen functional-based parametrized effective Hamiltonian,
we perform heating-up and cooling-down MD simulations.
In Fig.~\ref{fig:conpare-p},
lattice parameters as functions of temperature are plotted
under (a)~$p=0.0$~GPa, (b)~$p=-2.0$~GPa,
and (c)~$p=-0.005T$~GPa, where $T$ is the temperature in Kelvin.
$p=-2.0$~GPa and $p=-0.005T$ are
used to obtain a
lattice constant of 4.010~\AA\ just above $T_{\rm C}=403$~K.
Note that, at $T=400$~K, $p=-0.005T=-2.0$~GPa.
The latter temperature-dependent effective negative pressure
is for simulating thermal expansion.\cite{Leung:PhysRevB.65.214111.PZT,Silvia.PhysRevB.67.064106}
We note that adjustment of the lattice constant through a negative pressure(s) results
in significant improvement in the temperature of transition from cubic to tetragonal phase,
while the description of the lower two transitions improves only slightly.
Results with Wu-Cohen functional (in all the three schemes, $0.0$, $-2.0$, and $-0.005T$~GPa pressures)
show a significant improvements in cubic-to-tetragonal transition temperature
compared to the LDA-based results of previous MC~\cite{Zhong:V:R:PRB:v52:p6301:1995}
and MD~\cite{Nishimatsu:feram:PRB2008} calculations,
but not in the lower two transitions.
These mal-improvements in the lower two transitions may be coming from
difficulties in accurate first-principles calculations and polynomial fittings
of almost degenerated bottom of double wells of $[001]$, $[110]$, and $[111]$ distortions.
In Table~\ref{tab:Tc},
simulated transition temperatures are compared to
the previous MD simulations\cite{Nishimatsu:feram:PRB2008} with LDA-based parameters
and experimentally observed values.
Comparing them to experimentally observed temperature dependence of lattice parameters in
Ref.~\onlinecite{Kay:Vousden:Philosophical.Magazine:40:p1019-1040:1949},
we also note that $p=-0.005T$~GPa gives better temperature dependence of $c/a$ of tetragonal phase
than $p=0.0$ or $p=-2.0$~GPa,
though $c/a$ is still slightly overestimated.
\begin{table}
  \caption{Simulated
    cubic        $\leftrightarrow$ tetragonal,
    tetragonal   $\leftrightarrow$ orthorhombic, and
    orthorhombic $\leftrightarrow$ rhombohedral transition temperatures
    are compared to
    the previous MD simulations\cite{Nishimatsu:feram:PRB2008} with LDA-based parameters
    and experimentally observed values.
    Heating-up and cooling-down transition temperatures are averaged
    when corresponding transition has temperature hysteresis.}
  \label{tab:Tc}
  \centering
  \begin{tabular}{lp{3.5em}p{3.5em}p{3.5em}}
    \hline
    \hline
    XC functional and          & ortho. & tetra. & cubic \\
    effective negative pressure& rhombo.& ortho. & tetra.\\
    \hline
    GGA (W\&C), $0.0$ GPa      & 102 K  & 160 K & 288 K \\
    GGA (W\&C), $-2.0$ GPa     & 117 K  & 218 K & 408 K \\
    GGA (W\&C), $-0.005T$ GPa  & 103 K  & 187 K & 411 K \\
    \hline
    LDA, $ 0.0$ GPa            & ~95 K  & 110 K & 137 K \\
    LDA, $-5.0$ GPa            & 210 K  & 245 K & 320 K \\
    \hline
    experiment (after Refs. \onlinecite{Kay:Vousden:Philosophical.Magazine:40:p1019-1040:1949} and
                            \onlinecite{JOHNSON:APL:7:p221:1965})
                               & 183 K  & 278 K & 403 K \\
    \hline
    \hline
  \end{tabular}
\end{table}

\begin{figure}
  \centering
  \includegraphics[width=72mm]{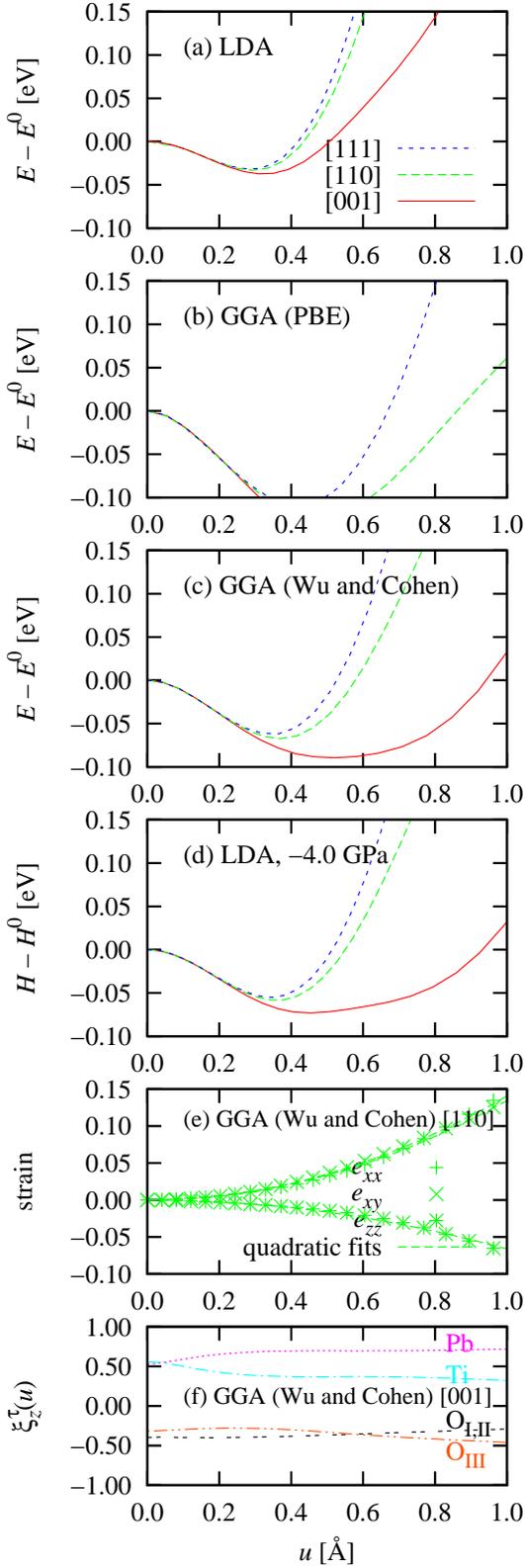}
  \caption{(Color online)
    (a)--(d) Total-energy surfaces for zone-center distortions of PbTiO$_3$.
    (e) $u$-dependence of strain along $[110]$ distortion.
    (f) ``Direction'' $\bm{\xi}(\bm{u})$ of atomic displacements along $[001]$ distortion.}
  \label{fig:comp-pto}
\end{figure}

\begin{figure}
  \centering
  \includegraphics[width=72mm]{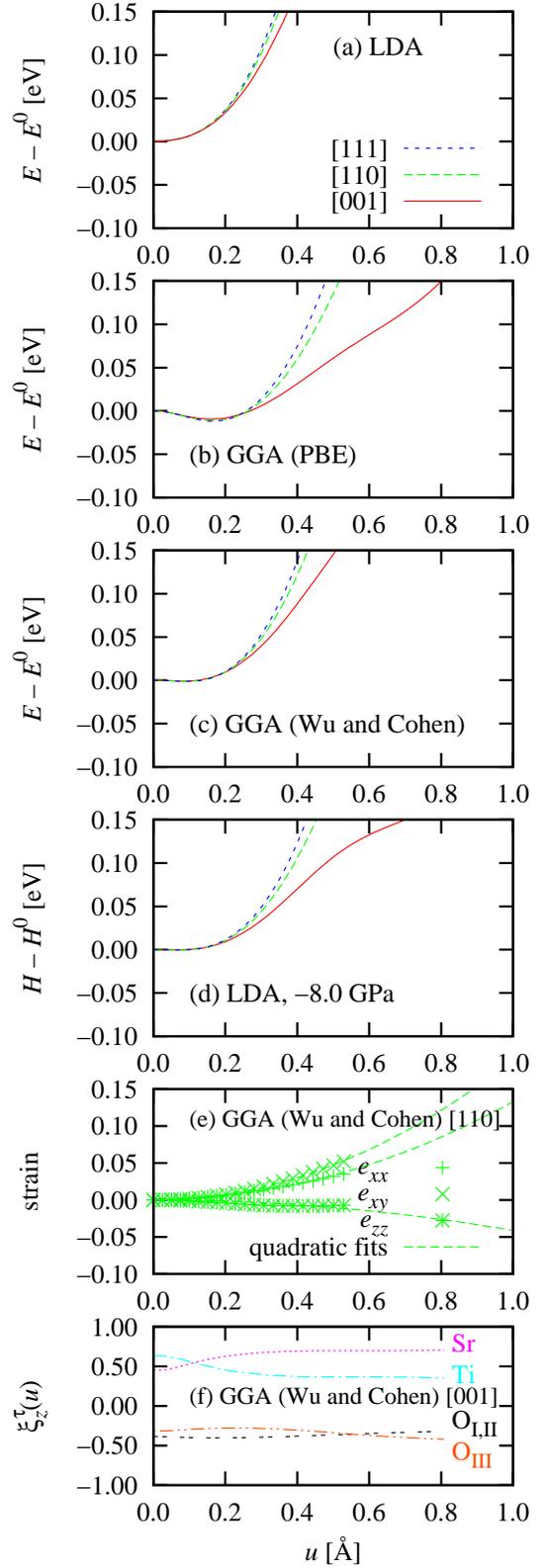}
  \caption{(Color online)
    (a)--(d) Total-energy surfaces for zone-center distortions of SrTiO$_3$.
    (e) $u$-dependence of strain along $[110]$ distortion.
    (f) ``Direction'' $\bm{\xi}(\bm{u})$ of atomic displacements along $[001]$ distortion.}
  \label{fig:comp-sto}
\end{figure}

We also calculated total-energy surfaces for zone-center distortions of PbTiO$_3$ and SrTiO$_3$
in Figs.~\ref{fig:comp-pto} and \ref{fig:comp-sto}, respectively.
Note that SrTiO$_3$ is not a ferroelectric material and
the polarizing zone-center distortion is not to be realized.
However, these may be useful data for investigating epitaxial constraint SrTiO$_3$ films
where the polarization properties depend crucially on the lattice mismatch.
In both cases,
similar trends in energetics as those in BaTiO$_3$ can be seen:
LDA results in shallow double wells and
GGA~(PBE) results in rather deep wells, while
GGA~(Wu and Cohen) results in intermediate depths of the double well potentials.
Again, LDA calculations under negative pressures,
$-4.0$~GPa for PbTiO$_3$ and
$-8.0$~GPa for SrTiO$_3$, give similar results to those of GGA~(Wu and Cohen).
Equilibrium cubic lattice constants are compared in Table~\ref{tab:a0}.

\section{Summary}
\label{sec:summary}
In this work, we have evaluated the improvement in the description of energy surface at 0~K relevant to
ferroelectricity in perovskite-based titanates with use of Wu-Cohen GGA functional for exchange
correlation energy in first-principles density functional theoretical calculations
to correctly estimate lattice constants and crystallographic anisotropy such as $c/a$ ratio.
We have demonstrated that the new GGA~(Wu and Cohen) functional based calculations
and LDA calculations under certain negative pressures
are capable of yielding fairly accurate and comparable total-energy surfaces of zone-center distortions
for $AB$O$_3$ perovskite-type ferroelectrics; BaTiO$_3$, PbTiO$_3$, and SrTiO$_3$.
We have shown that their polar structural distortions are highly sensitive to their lattice constants,
and hence much of the improvement with Wu-Cohen functional comes from its ability to correctly estimate
lattice constants.
We note that the use of Wu-Cohen functional has
little influence elastic parameters, the soft mode eigenvectors and
mode effective charges which govern the long-range dipolar interactions.
What is most affected are the terms of the cubic anisotropy, as
reflected in the strain coupling term $B_{1yy}$ and anharmoinc terms,
and hence in the relative energy well-depths of polar distortions along
$[001]$, $[110]$, and $[111]$ directions. This is not quite surprising as
the motivation in development of the Wu-Cohen functional was to get
structural properties such as $c/a$ ratio and lattice constants with
better accuracy.

We then analyzed consequences of this improvement in energy functional to finite temperature
ferroelectric transition, taking an example of BaTiO$_3$. To this end, starting
from calculations with GGA~(Wu and Cohen) functional,
we constructed a new parameter set for effective Hamiltonian of BaTiO$_3$ employing the valley-tracing
technique that effectively includes anharmonic coupling of the soft polar mode with higher energy
polar modes. Comparing this and an LDA-based effective Hamiltonian with MD simulations,
we find that the use of Wu-Cohen functional leads to a clear improvement in description of the highest
temperature transition from cubic to tetragonal phase.
We also confirmed that,
as already mentioned in Refs. \onlinecite{Leung:PhysRevB.65.214111.PZT} and \onlinecite{Silvia.PhysRevB.67.064106},
the effect of thermal expansion, which is basically coming from the odd
order energy terms of atomic displacements and their coupling with strains, cannot be ignored as these
materials exhibit strong electro-mechanical couplings. Secondly, accounting for
thermal expansion approximately through the temperature-dependent effective negative pressure in
effective Hamiltonian, a
more realistic description of ferroelectric phase transition can be obtained.

\section*{Acknowledgments}
Computational resources
were provided by the Center for Computational Materials Science,
Institute for Materials Research (CCMS-IMR), Tohoku University.
We thank the staff at CCMS-IMR for their constant effort.
This study was also supported by the Next Generation Super Computing Project,
Nanoscience Program, MEXT, Japan.
UVW acknowledges an IBM faculty award grant in supporting some of his work.

\bibliography{biblio/ferroelectrics,biblio/phonon,biblio/abinitio,biblio/EPAPS,biblio/ElasticModuli,biblio/MD}

\end{document}